\documentclass[aps,prb,superscriptaddress,tightenlines,twocolumn,floatfix,notitlepage]{revtex4-1}
\usepackage{graphicx}
\usepackage[english]{babel}
\usepackage{amsmath}
\usepackage{amssymb}
\usepackage{tensor}

\newcommand{\vk}{\mathbf{k}}
\newcommand{\vn}{\mathbf{n}}
\newcommand{\vm}{\mathbf{m}}
\newcommand{\be}{\begin{eqnarray}}
\newcommand{\ee}{\end{eqnarray}}

\newcommand{\hx}{\hat{x}}
\newcommand{\hy}{\hat{y}}
\newcommand{\dc}{c^{\dagger}}

\def\ket#1{|#1\rangle}
\def\bra#1{\langle #1 |}
\def\ep#1{\langle #1 \rangle}

\begin{document}

\title{Dynamics of 2D topological quadrupole insulator and Chern insulator induced by real-space topological changes}

\author{Yan He}
\affiliation{College of Physics, Sichuan University, Chengdu, Sichuan 610064, China}
\email{heyan$_$ctp@scu.edu.cn}

\author{Chih-Chun Chien}
\affiliation{School of Natural Sciences, University of California, Merced, CA 95343, USA.}
\email{cchien5@ucmerced.edu}

\begin{abstract}
The dynamics of two-dimensional (2D) topological quadrupole insulator (TQI) and Chern insulator (CI) after the real-space configuration is transformed from a cylinder or Mobius strip to open boundary condition (OBC) and vice versa is analyzed. Similar dynamics of both models are observed, but the quadrupole corner states of the TQI makes the signatures more prominent. After the systems transform from a cylinder or Mobius strip to OBC, the occupation of the corner state of the TQI and the edge state of the CI exhibits steady-state behavior. The steady-state values depend on the ramping rate of the configuration transformation, manifesting a type of quantum memory effect. On the other hand, oscillatory density ripples from the merging of edge states persist after the systems transform from OBC to a cylinder or Mobius strip. If the final configuration is a cylinder, the density ripples are along the edges of the cylinder. In contrast, the density ripples can traverse the bulk after the systems transform from OBC to a Mobius strip. The transformation of real-space topology thus can be inferred from the dynamical signatures of the topological edge states.
\end{abstract}

\maketitle

\section{Introduction}
The discovery of topological insulators has revolutionized our classification of electronic systems and accelerated the search for novel materials~\cite{Kane_TIRev,Zhang_TIRev,ShenTI,Asboth2016,Chiu2016,Stanescu_book}. The bulk-boundary correspondence connects the quantized topological invariants defined in systems with periodic boundary condition to the numbers of edge states in systems with open boundary condition. Recently, a class of topological insulators called the higher-order topological insulator (HOTI)~\cite{Bernevig,Schindler18} has been shown to have hinge or corner states with codimensions larger than one. The mechanism of higher-order topological insulators transcends material properties. In addition to electronic materials~\cite{HOTI_Bi}, HOTIs have been realized in photonic~\cite{Mittal19}, mechanical~\cite{Matlack18}, phononic~\cite{Garcia18,Huo19}, acoustic~\cite{LinSonic19,Xue19} systems and electric-circuit simulators~\cite{Imhof18}. Moreover, possible higher-order topological superconductors have be discussed~\cite{HOSC18,Volpez19} and HOTI may also be found in twisted bilayer graphene~\cite{Park19}.

While the topology of the band structure gives rise to nontrivial transport properties~\cite{Kane_TIRev,Zhang_TIRev,ShenTI,Asboth2016,Chiu2016,Stanescu_book}, the topology in real space may also lead to interesting phenomena. One simple yet topologically nontrivial manifold is the Mobius strip, compared to the topologically trivial cylinder~\cite{Nakahara_book}. There have been studies of static properties of topological systems on a Mobius strip, showing nontrivial states across the bulk that is not possible in the cylinder configuration~\cite{Mobius11,Mobius14}. On the other hand, dynamics of topological systems after a global change in the parameters~\cite{Dauphin13,Sacramento14,Hauke14,Caio15,Wang16,Wang17} or a local change of boundary condition~\cite{He2016a} can show signatures of the topological states in time evolution. Here we present our study of the dynamics of two-dimensional (2D) topological quadrupole insulator (TQI) and Chern insulator (CI) as the real-space configuration changes from a rectangle with open boundary condition (OBC) to a cylinder with periodic boundary condition along one direction or a Mobius strip with twisted boundary condition along one direction, and vice versa. The TQI is an example of the HOTI supporting a quadrupole moment in the presence of OBC while the CI has chiral topological edge states at the boundary. The 2D cases studied here exhibit richer physics compared to the 1D cases shown in Ref.~\cite{He2016a} because (1) the HOTIs require the codimension of edge states to be greater than one, and (2) twisted real-space manifolds, such as the Mobius strip, are realizable in higher dimensions.

Although the dynamics of the 2D TQI and CI show many similarities as the boundary changes, the quadrupole corner modes of the TQI make the signatures more prominent. Thus, we present the results of TQI first and then compare with the CI. For the 2D TQI from a cylinder or Mobius strip to OBC, the initial condition decays away and the system develops a quadrupole moment in both scenarios. Interestingly, we have not introduced any dissipation mechanism, but the system reaches a steady state with the density distribution becoming time-independent after the transient behavior dies out. Interestingly, the steady-state occupation of the quadrupole corner state exhibits an explicit dependence on the rate of the boundary-condition change. The slower the ramping rate, the higher the occupation of the corner state can be observed. This type of behavior is a rate-dependent quantum memory effect~\cite{LaiMemory}, which has been discussed in the dynamics of 1D topological models as the boundary condition changes from periodic to open~\cite{He2016a}. Moreover, the dynamics of the quadrupole moment exhibits similar quantum memory effect, although the fluctuation of the steady-state value is more significant. While the 2D CI does not have a finite quadrupole moment, it has chiral topological edge states at the boundary. The occupation of the edge state of the 2D CI shows similar rate-dependent memory effect after the system transforms from a cylinder or Mobius strip to OBC.

As the systems transform from OBC to a cylinder or Mobius strip, there is no longer a steady state because the systems exhibit oscillatory density ripples for both the TQI and CI. For the 2D TQI, the density ripples in the final cylinder configuration only oscillates along the two edges of the cylinder. In contrast, the density ripples in the Mobius strip only oscillates along a line perpendicular to its edge. As pointed out in Refs.~\cite{Mobius11,Mobius14}, the Mobius strip can frustrate the would-be topological edge state and cause it to traverse the bulk instead. The different dynamics of the density ripples from OBC to a cylinder and Mobius strip clearly shows how a change in the real-space topology can drive the topological edge states and allows observable effects out of equilibrium. On the other hand, the topological edge state of the 2D CI leads to a weaker contrast between the cylinder and Mobius strip configurations. After a transformation from OBC to a cylinder, the density ripples of the 2D CI are still confined in the two edges of the cylinder. However, the density ripples of the CI from OBC to a Mobius strip oscillates in both directions, with the amplitude in the direction perpendicular the edge dominates.

The rest of the paper is organized as follows. Sec.~\ref{sec:theory} summarizes the TQI and CI models in different configurations. The time-evolution and initial state are also explained. Sec.~\ref{sec:numerical} presents the dynamics of the systems when the real-space configurations are transformed. Features of quantum memory effect and density patterns are analyzed, and the similarities and differences between the TQI and CI models are summarized. Sec.~\ref{sec:exp} discusses some implications for experiments and possible realizations. Sec.~\ref{sec:conclusion} concludes our work.

\section{Theoretical framework}\label{sec:theory}
\subsection{Topological quadrupole insulator and its real-space configurations}
We consider a 2D TQI model given by the following Hamiltonian of a four-band model in $k$-space~\cite{Bernevig}:
\be\label{eq:TQI_Hk}
&&H=\sum_{\vk,a,b} \psi_{\vk,a}^{\dagger}h_{ab}(\vk)\psi_{\vk,b},\\
&&h(\vk)=(m+w\cos k_x)\Gamma_4+w\sin k_x\Gamma_3\nonumber\\
&&\qquad+(m+w\cos k_y)\Gamma_2+w\sin k_y\Gamma_1+\delta\Gamma_0.
\ee
Here $\psi_{\vk}=(c_{\vk,1},c_{\vk,2},c_{\vk,3},c_{\vk,4})^T$ with the superscript $T$ denoting the transpose, $\Gamma_4=\sigma_1\tau_0$, $\Gamma_i=\sigma_2\tau_i$ with $i=1,2,3$, and $\Gamma_0=\sigma_3\tau_0$. $\sigma_i$ and $\tau_i$ with $i=1,2,3$ are the Pauli matrices and $\sigma_0$, $\tau_0$ are the 2 by 2 identity matrices in different internal spaces. $m$ gives the onsite potentials and $w$ is the hopping coefficient. This model may be viewed as a 2D generalization of the Su-Schrieffer-Heeger model~\cite{SSH79}. The topological nontrivial phase arises when $|m|<|w|$ and has a quantized electric quadrupole moment $Q_{xy}$. There are two reflection symmetries given by the operators $\hat{m}_x=\sigma_1\tau_3$ and $\hat{m}_y=\sigma_1\tau_1$, whose operations on the Hamiltonian are
\be
\hat{m}_x h(k_x,k_y)\hat{m}_x^{\dagger}=h(-k_x,k_y),\\
\hat{m}_y h(k_x,k_y)\hat{m}_y^{\dagger}=h(k_x,-k_y).
\ee
Those symmetries quantize both components of the polarization and also the quadrupole moment. The term $\delta\Gamma_0$ breaks those two reflection symmetries, so the quadrupole moment is not quantized exactly. In the following, we will assume $\delta<w$ and use it to fix the sign of $Q_{xy}$.

In real space, the model is given by
\be
&&H=\sum_{\vn}\Big[\psi_{\vn}^{\dagger}\Big(m(\Gamma_4+\Gamma_2)+\delta\Gamma_0\Big)\psi_{\vn}\nonumber\\
&&+\frac{w_{\vn}}2\psi_{\vn}^{\dagger}(\Gamma_4-i\Gamma_3)\psi_{\vn+\hx}
+\frac{w}2\psi_{\vn}^{\dagger}(\Gamma_2-i\Gamma_1)\psi_{\vn+\hy}\nonumber\\
&&+\frac{w_{\vn}}2\psi_{\vn+\hx}^{\dagger}(\Gamma_4+i\Gamma_3)\psi_{\vn}
+\frac{w}2\psi_{\vn+\hy}^{\dagger}(\Gamma_2+i\Gamma_1)\psi_{\vn}\Big].
\label{Hxy}
\ee
Here $\vn=(n_x,n_y)$ is the coordinate of a 2D square lattice with $n_x,n_y\in\{1,\cdots,N\}$. $\hx$ and $\hy$ are the unit vectors along the $x$ and $y$ axes, respectively. We introduce $\psi_{\vn}=(c_{\vn,1},\cdots,c_{\vn,4})^T$. After solving the model with open boundary condition along both $x$ and $y$ axes, one will find four corner states localized at the corners of the rectangle. Unlike the SSH model, where the number of edge states depends on the parity of the total number of sites, the 2D TQI model has the same number of corner states regardless of the parity of $N$.

To study the dynamics induced by a change of the real-space topology, we consider the following configurations: (1) A rectangle with open boundary condition (OBC) in both $x$ and $y$ directions. (2) A cylinder with periodic boundary condition (PBC) along one direction (say, the $x$ direction) and open boundary condition along the other direction (say, the $y$ direction). (3) A Mobius strip with twisted boundary condition (TBC) along the $x$-direction and open boundary condition along the $y$ direction. The periodic (or twisted) boundary condition along the $x$ axis leads to $\psi_{N+1,i}=\psi_{1,i}$ (or $\psi_{N+1,i}=\psi_{1,N-i+1}$). Since periodic or twisted boundary condition works if $N$ is even, in the following we will focus on the configurations with even $N$.

The transformation from configuration (1) to configuration (2) or (3) or vice versa can be achieved by tuning the hopping coefficient $w_{\vn}$ associated with the link between site $\vn$ and site $\vn+\hx$. If we consider a linear transformation with a constant ramping rate, the system will evolve from configuration (2) to configuration (1) when the hopping coefficients transform according to
\be
w_{\vn}=\left\{
      \begin{array}{ll}
        w, & n_x\neq N; \\
        w(t), & n_x=N.
      \end{array}
    \right.
\ee
Here $w(t)=w(1-t/t_q)$ for $0\le t\le t_q$ and $w(t)=0$ for $t>t_q$. In the following we will focus on linear transformations between different configurations.

\subsection{Time evolution and initial condition of TQI}
The quantum dynamics can be describe by the Heisenberg equation~\cite{He2016a}.
For an operator $A$, we have $\frac{d A}{d t}=-i[A,H]$.
Here we set $\hbar=1$. The time unit is $t_0=\hbar/w$.
We define the following correlation functions as a set of 4 by 4 matrices:
\be\label{eq:Gnm}
G(\vn,\vm)=\bra{S_0}\psi_{\vn}\psi^{\dagger}_{\vm}\ket{S_0}.
\ee
Here $\ket{S_0}$ is the initial quantum state. Once the time-evolution of the correlation functions are obtained, physical quantities such as the density profile can be found. The detailed expressions of the time-evolution equations are summarized in App.~\ref{sec:Evo_TQI}.

The initial condition of the OBC, cylinder, or Mobius strip corresponds to the half-filled ground state of the corresponding initial Hamiltonian. The details of the initial conditions are summarized in App.~\ref{sec:Ini_TQI}.

\subsection{Time evolution and initial condition of 2D Chern insulator}
To contrast the influence of the quadruple corner state of the TQI, we also consider the time evolution of a 2D two-band Chern insulator (CI) model with the Hamiltonian
\be
H_{CI}&=&w\sin k_x\sigma_1+w\sin k_y\sigma_2 \nonumber \\
& &+(m+w\cos k_x+w\cos k_y)\sigma_3.
\ee
For $0<m<2$, the lower band of the model has Chern number $C_1=1$. In real space, the Hamiltonian can be rewritten as
\be
&&H_{CI}=\sum_{\vn}\Big[m\psi_{\vn}^{\dagger}\sigma_3\psi_{\vn}\nonumber\\
&&+\frac{w_{\vn}}{2}\psi_{\vn}^{\dagger}(\sigma_3-i\sigma_1)\psi_{\vn+\hx}
+\frac{w}{2}\psi_{\vn}^{\dagger}(\sigma_3-i\sigma_2)\psi_{\vn+\hy}\nonumber\\
&&+\frac{w_{\vn}}{2}\psi_{\vn+\hx}^{\dagger}(\sigma_3+i\sigma_1)\psi_{\vn}
+\frac{w}{2}\psi_{\vn+\hy}^{\dagger}(\sigma_3+i\sigma_2)\psi_{\vn}
\Big].
\label{eq-CI}
\ee
Here $\psi_{\vn}=(c_{1,\vn},\,c_{2,\vn})^T$, and one needs to impose proper boundary condition.

Similar to the analysis of the TQI, we define the correlation functions as a set of 2 by 2 matrices:
\be
G(\vn,\vm)=\bra{S_0}\psi_{\vn}\psi^{\dagger}_{\vm}\ket{S_0}.
\ee
Here $\ket{S_0}$ is the initial quantum state. The time-evolution equations of the correlation functions are summarized in App.~\ref{sec:Evo_CI}. Physical quantities such as the density profiles can be obtained once the time-evolved correlation functions are found. The initial condition corresponds to the half-filled ground state of the initial Hamiltonian in the OBC, cylinder, or Mobius strip configuration. The details of constructing the initial conditions are also summarized in App.~\ref{sec:Evo_CI}.

\section{Numerical Results}\label{sec:numerical}
\subsection{Dynamics of 2D TQI}

\begin{figure*}
	\includegraphics[width=\textwidth]{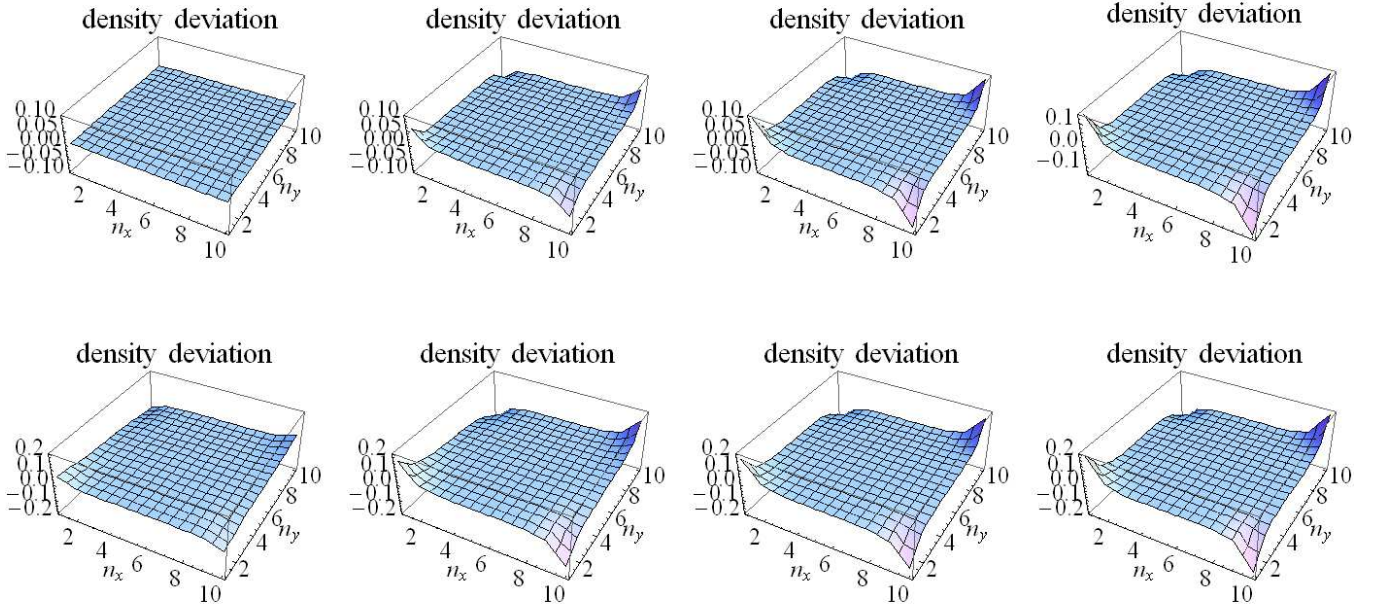}
	\caption{Snapshots of the density deviation from the average density, $\Delta\rho(\vn)$, of the 2D TQI model at $t/t_0=1,5,7,10$ (from the left to the right). The evolution from a cylinder (or a Mobius strip) to OBC is shown in the upper (or lower) row. Here $t_q/t_0=10$, $w=1$, $m=0.5$, $\delta=0.2$, and $N=10$. The initial ground state is half filled. }
	\label{fig:TQI}
\end{figure*}

\begin{figure}
	\centerline{\includegraphics[width=\columnwidth]{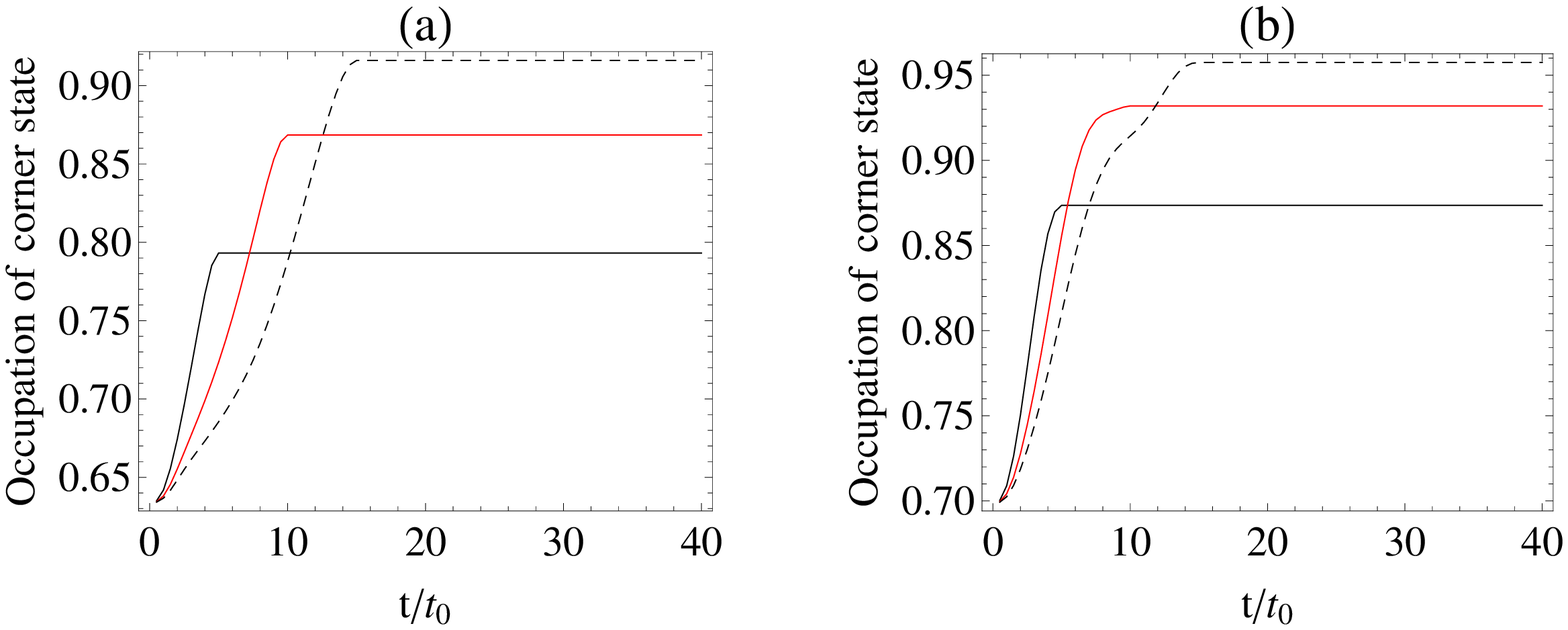}}
	\caption{Time evolution of the corner-state occupation of the TQI model from a cylinder to OBC (a) and from a Mobius strip to OBC (b). The black solid lines, red solid lines, and black dashed line correspond to $t_q/t_0=5,10,15$, respectively. The different steady-state values reflect memory effect of the ramping rate. Here $w=1$, $m=0.5$, $\delta=0.2$, and $N=10$. The initial ground state is half filled. }
	\label{corner}
\end{figure}

\begin{figure}
\centerline{\includegraphics[width=\columnwidth]{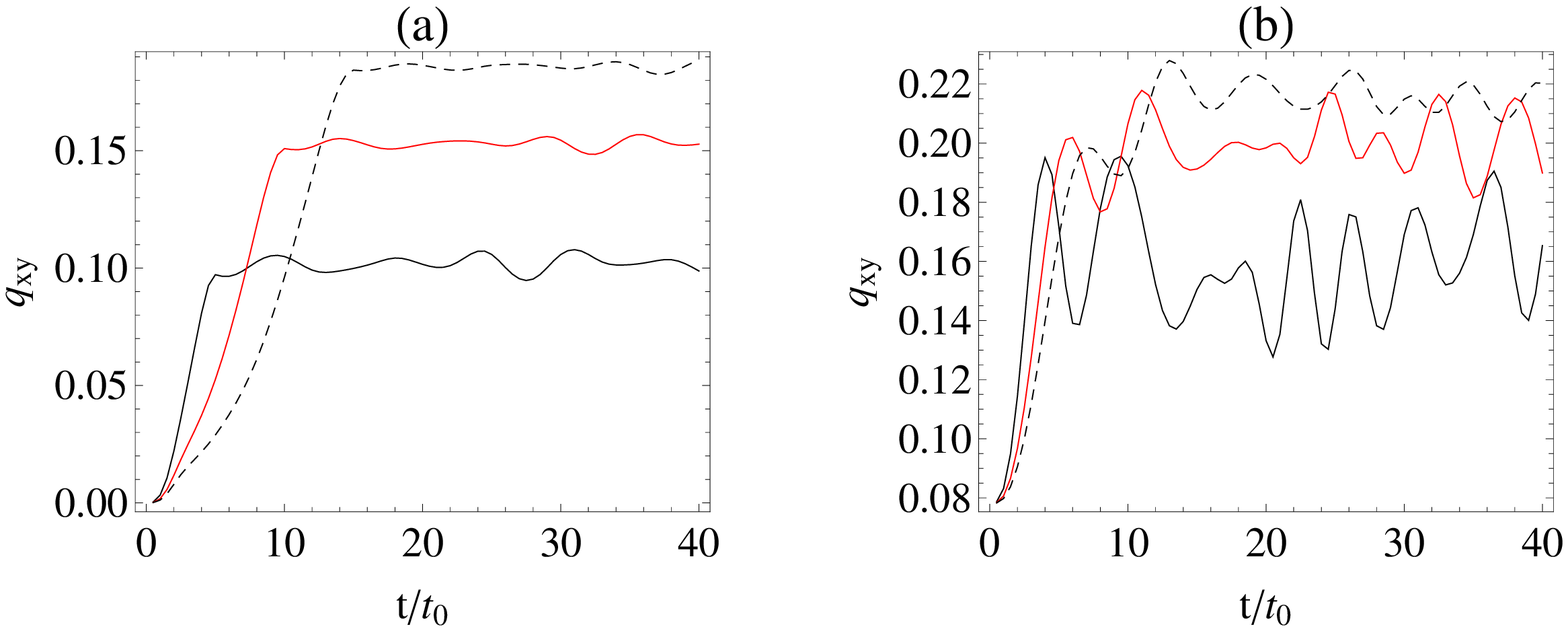}}
\caption{Time evolution of the physical quadrupole moment $q_{xy}$ of the TQI model defined in Eq.~\eqref{eq:qxy}. (a) and (b) show the cases from a cylinder to OBC and from a Mobius strip to OBC, respectively. The black solid lines, red solid lines, and black dashed lines correspond to $t_q/t_0=5,10,15$, respectively. Here $w=1$, $m=0.5$, $\delta=0.2$, and $N=10$. The initial ground state is half filled.}
\label{qxy}
\end{figure}

\begin{figure*}
	\includegraphics[width=\textwidth]{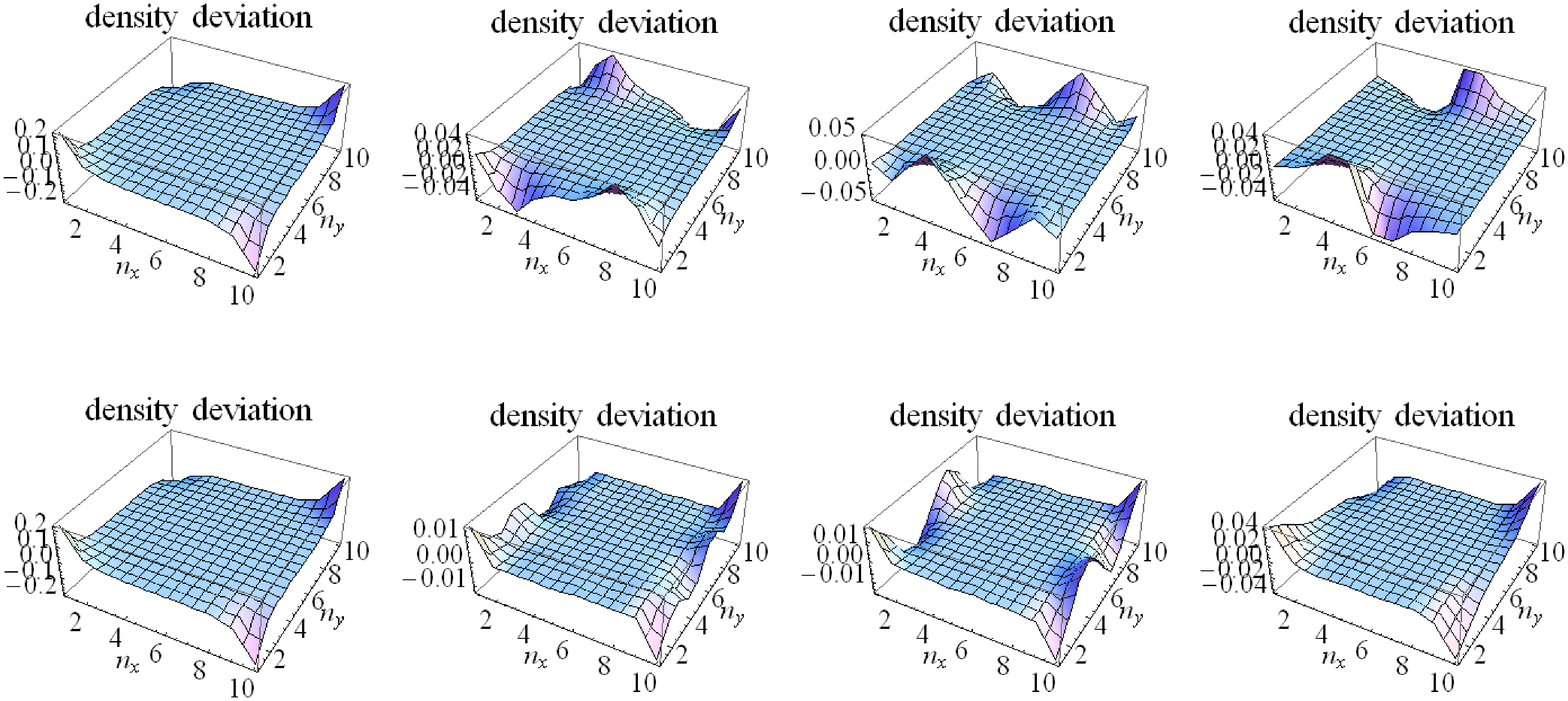}
	\caption{Snapshots at $t/t_0=10,12,15,20$ (from left to right) of the density deviation from the average density of the TQI model for the evolution from OBC to a cylinder (top row) and from OBC to a Mobius strip (bottom row). Here $t_q/t_0=10$, $w=1$, $m=0.5$, $\delta=0.2$, and $N=10$. The initial state is the ground state of the initial Hamiltonian with half-filling.}
	\label{dens-evolve}
\end{figure*}

\subsubsection{From cylinder or Mobius strip to OBC}
In Figure \ref{fig:TQI}, we present the time evolution of the density deviation from the average density of the TQI model for the two cases where the configuration evolves from a cylinder or Mobius strip to OBC. We choose the energy unit so that $w=1$ and present the results with $m=0.5$ and $\delta=0.2$. We have verified that the conclusions remain qualitatively the same for different values of $\delta >0$. The initial state is obtained with PBC or TBC along the $x$-axis. The hopping coefficients at the emerging boundary along the $x$-direction take the form $w_{\vn}(t)=w(1-t/t_q)$ for $0\le t\le t_q$ and $w_{\vn}=0$ for $t>t_q$. We show the results with ramping time $t_q/t_0=10$ first and will analyze the effect of tuning $t_q$ shortly.

In our simulations, a square lattice labeled by $n_x,\,n_y=1,\cdots, N$ with $N=10$ is used. The size is large enough that in the static configuration with OBC, the quadrupole corner states decay to negligible values away from the corners. The density at each site is the sum of the densities of the fermions from the four sublattices. Explicitly,
\be
\rho(\vn)=\sum_{i=1}^4\ep{\dc_{i,\vn}c_{i,\vn}}=\sum_{i=1}^4[1-G_{ii}(\vn,\vn)].
\ee
The initial state is the half-filled ground state of fermions. To better present the results, we subtract the average density from the results and show $\Delta\rho(\vn)\equiv\rho(\vn)-2$ in the figures.

The top row of Fig.~\ref{fig:TQI} shows the time evolution of density deviation as the configuration changes from a cylinder to OBC. The quadrupole moment can be inferred from the two peaks and two dips at the four corners. As the links connecting $n_x=1$ and $n_x=10$ decay, the quadrupole moment starts to grow. However, the growth stops after the transformation is completed, and the system reaches a steady state. The emergence of a steady state is interesting because we do not introduce any explicit dissipation mechanism. Steady states of systems without dissipation have also been found in the dynamics of 1D topological systems after a change in boundary condition~\cite{He2016a}.

The time evolution of the TQI model from a Mobius strip to OBC along the $x$-direction is studied in a similar way. We use the same parameters as the case from a cylinder to OBC and show the time evolution of the density deviation in the bottom row of Fig.~\ref{fig:TQI}. The results are qualitatively similar to the case from a cylinder to OBC shown in the top row. Again, the system enters a steady state after the transformation is completed in the absence of any dissipation.

While the dynamics of the cases from a cylinder to OBC and from a Mobius strip to OBC are similar, they both show interesting quantum memory effect. To characterize the behavior, we note that the final OBC configurations of the two cases are the same and there are four in-gap corner states. Due to the term $\delta\Gamma_0$ of Eq.~\eqref{eq:TQI_Hk}, two of them are below the zero-energy and the other two are above. For the static TQI with OBC at half-filling, the two lower-energy corner states are filled while the two higher-energy corner states are empty.

We found that the four corner states exhibit the same dynamics in the sense that (1) the dynamics of the two lower-energy corner states are identical up to numerical precision and (2) if the occupation of the low-energy corner state is $\alpha(t)$, that of the higher-energy corner state is $1-\alpha(t)$.  Thus, we pick one low-energy corner state and define its operator by $\eta_c=\phi_{i,c}^* c_i$, which then gives the density
\be
\ep{\eta_c^{\dagger}\eta_c}=1-\ep{\eta_c\eta_c^{\dagger}}=1-\sum_{i,j}\phi_{i,c}^*\ep{c_i\dc_j}\phi_{j,c}.
\ee
In Figure \ref{corner}, we show the time evolution of the occupation of the selected corner state. The black solid, red solid and black dashed lines correspond to $t_q/t_0=5$, $10$, and $15$, respectively. The left (right) panel corresponds to the case from a cylinder to OBC (from a Mobius strip to OBC).

As one can see, the occupation of the corner state reaches a constant after the links at the $x$-direction boundary are completely turned off, signaling the emergence of a steady state even in absence of dissipation. Importantly, the steady-state values depend on the ramping rate of the boundary links. As the ramping rate increases (corresponding to decreasing $t_q$), the steady-state occupation of the corner state decreases accordingly. Similar quantum memory effects of topological edge modes in several 1D topological systems have been found in Refs.~\cite{He2016a}. The quantum memory effects come from the localized nature of the corner states, which trap particles once their wavefunctions overlap with the corner states. Since the excitation of particles depends on the ramping rate of the transformation, the occupation of the corner state thus depends on the ramping rate. In the adiabatic limit ($t_q\rightarrow\infty$), the two lower-energy corner states would be fully occupied for a half-filled system. For transformations with finite $t_q$, the excitation of quantum particles to other mobile quantum states reduces the occupation of the corner states, so they are not fully occupied in the steady state.

The quadrupole moment $Q_{xy}$ of the TQI with PBC comes from its band structure~\cite{Bernevig}. Here we define a physical quadrupole moment as
\be\label{eq:qxy}
q_{xy}=\frac{1}{N^2}\sum_{\vn}\Delta\rho(\vn)n_xn_y,
\ee
which has been normalized by the system size. Figure \ref{qxy} shows the time evolution of $q_{xy}$ as the system transforms from a cylinder or Mobius strip to OBC. The black solid, red solid, and black dashed lines correspond to $t_q/t_0=5$, $10$, and $15$, respectively. For the case from a cylinder to OBC, $q_{xy}$ reaches a steady-state value with relatively small fluctuations after $t_q$. The steady-state value depends on the ramping rate, so the physical quadrupole moment also exhibits quantum memory effect similar to that of the occupation of the corner state. For the case from a Mobius strip to OBC, however, $q_{xy}$ exhibits more significant fluctuations around the steady-state value after $t_q$. Although the steady-state values in this case also exhibit quantum memory effect, the fluctuations may make its observation more challenging. We remark that the quadrupole moment~\eqref{eq:qxy} receive contributions from all the states, not just from the corner state. The oscillatory contributions from the bulk states in~\eqref{eq:qxy} are responsible for the fluctuations shown in Fig.~\ref{qxy}.

\subsubsection{From OBC to cylinder or Mobius strip}
Next, we consider the transformations from OBC to a cylinder or Mobius strip by tuning the time-dependent hopping coefficients at the $x$-direction boundary. Explicitly, $w_{\vn}(t)=w t/t_q$ for $0\le t\le t_q$ and $w_{\vn}=w$ for $t>t_q$. The parameters are chosen to be the same as the previous cases from a cylinder or Mobius strip to OBC. Figure \ref{dens-evolve} shows the time evolution of the density deviation of the TQI model at time $t/t_0=10,12,13,20$ (from the left to the right). The top (bottom) row shows the results from OBC to a cylinder (Mobius strip). Importantly, the real-space topology differentiates the dynamics in the two cases: For the case from OBC to a cylinder, oscillatory density ripples are observable along the $x$-direction at the two edges of the cylinder. In contrast, for the case from OBC to a Mobius strip, oscillatory density ripples are observable along the $y$-direction, traversing the bulk of the Mobius strip.

The density ripples along the different directions can be understood as follows: The initial state has a quadrupole moment shown by the two peaks and two dips at the corners. When the system evolves from OBC to a cylinder, the two density peaks encounters the two density dips along the $x$-direction. Since we do not include any dissipation mechanism, the peaks and dips cannot annihilate each other completely, so they cause density ripples along the two edges in the $x$-direction. In contrast, when the system evolves from OBC to a Mobius strip, the two peaks meet each other while the two dips also meet each other and they form a dipole across the $y$ axis. The merged peak and dip then interact with each other, and they cause density ripples along the $y$ axis instead.

One explanation of the absence of density ripples in Fig.~\ref{fig:TQI} when the system evolves from a cylinder or Mobius strip to OBC is that the corner states emerge in the final configuration and being zero-energy localized states, the corner states collect its occupation while the bulk states sum up to the average density. In contrast, when the system evolve from OBC to a cylinder or Mobius strip, the corner states merge into the bulk states and acquires dispersion in their energy. The dynamic phase $e^{-iE_nt}$, where $E_n$ is the eigenenergy, leads to oscillatory behavior of the final configuration. Since the eigenenergy depends on the system parameters $(w,\delta)$ as well as the system size, the wavelength and period of the density ripples vary when the parameter or size changes. Nevertheless, the propagation direction of the density ripples is determined by the real-space topology, which is robust against changes of the parameter or size. 

In Ref.~\cite{Mobius11,Mobius14}, the static properties of some 2D topological models are studied with periodic or twisted boundary condition. It was found that some models on a Mobius strip may form states traversing the bulk because of the way the two sides of a rectangle is merged. Here we found the dynamics of the 2D TQI with a quadrupole moment can have density ripples traversing the bulk if the final configuration is a Mobius strip. Therefore, one can tell if the system transforms into a cylinder or a Mobius strip by monitoring the patterns of the density ripples.

\begin{figure*}
\includegraphics[width=\textwidth]{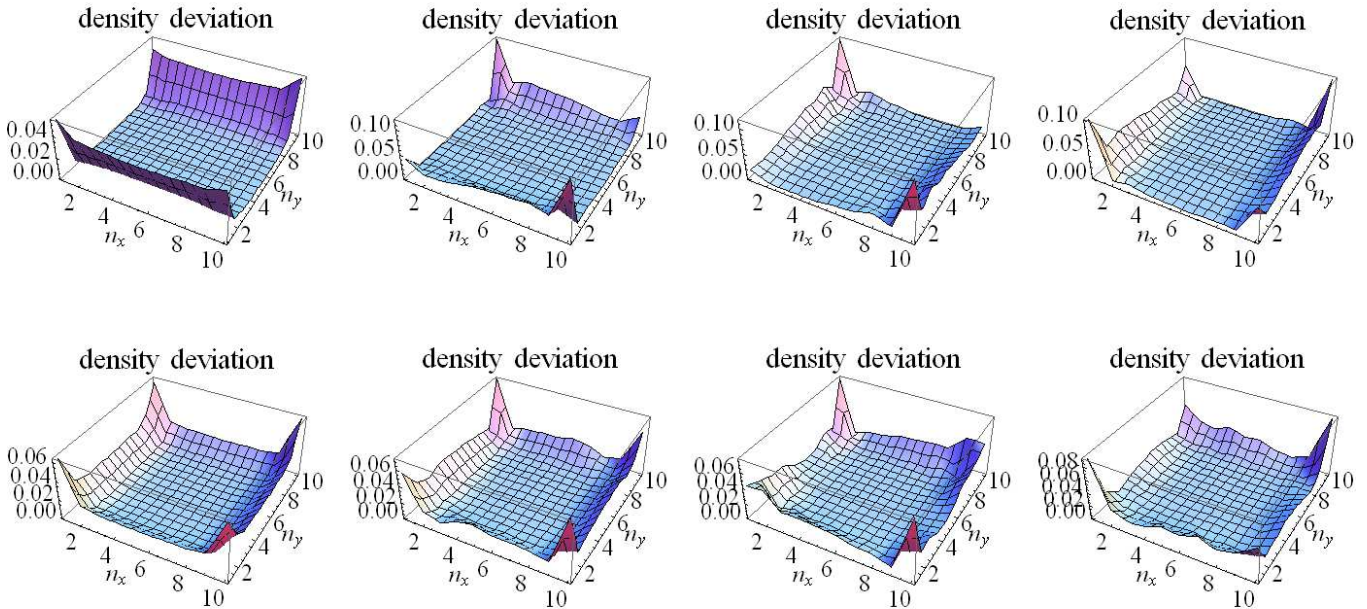}
\caption{Snapshots of the density deviation from the average density of the 2D two-band Chern insulator model at $t/t_0=1,8,16,20$ (from the left to the right). The case evolving from a cylinder (or a Mobius strip) to OBC is shown in the top row (or bottom row). Here $t_q/t_0=10$, $w=1$, $m=0.5$, and $N=10$. The initial ground state is half-filled with the edge states occupied.
}
\label{CI}
\end{figure*}

\begin{figure}
	\centerline{\includegraphics[width=\columnwidth]{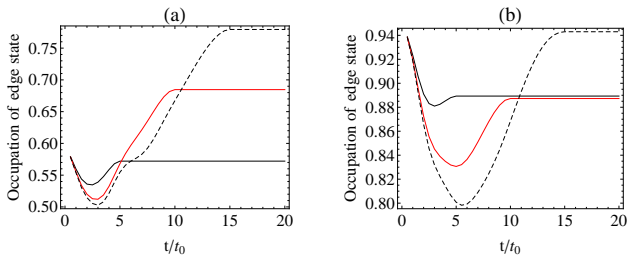}}
	\caption{Time evolution of the occupation of the topological edge state of the 2D two-band CI model from a cylinder to OBC (a) and from a Mobius strip to OBC (b) with selected ramping times $t_q/t_0=5$ (black lines), $t_q/t_0=10$ (red lines), and $t_q/t_0=15$ (black dashed lines). Here $w=1$, $m=0.5$, and $N=10$. The initial ground state is half-filled with the edge states occupied.}
	\label{fig:CI_edge}
\end{figure}

\begin{figure*}
\includegraphics[width=\textwidth]{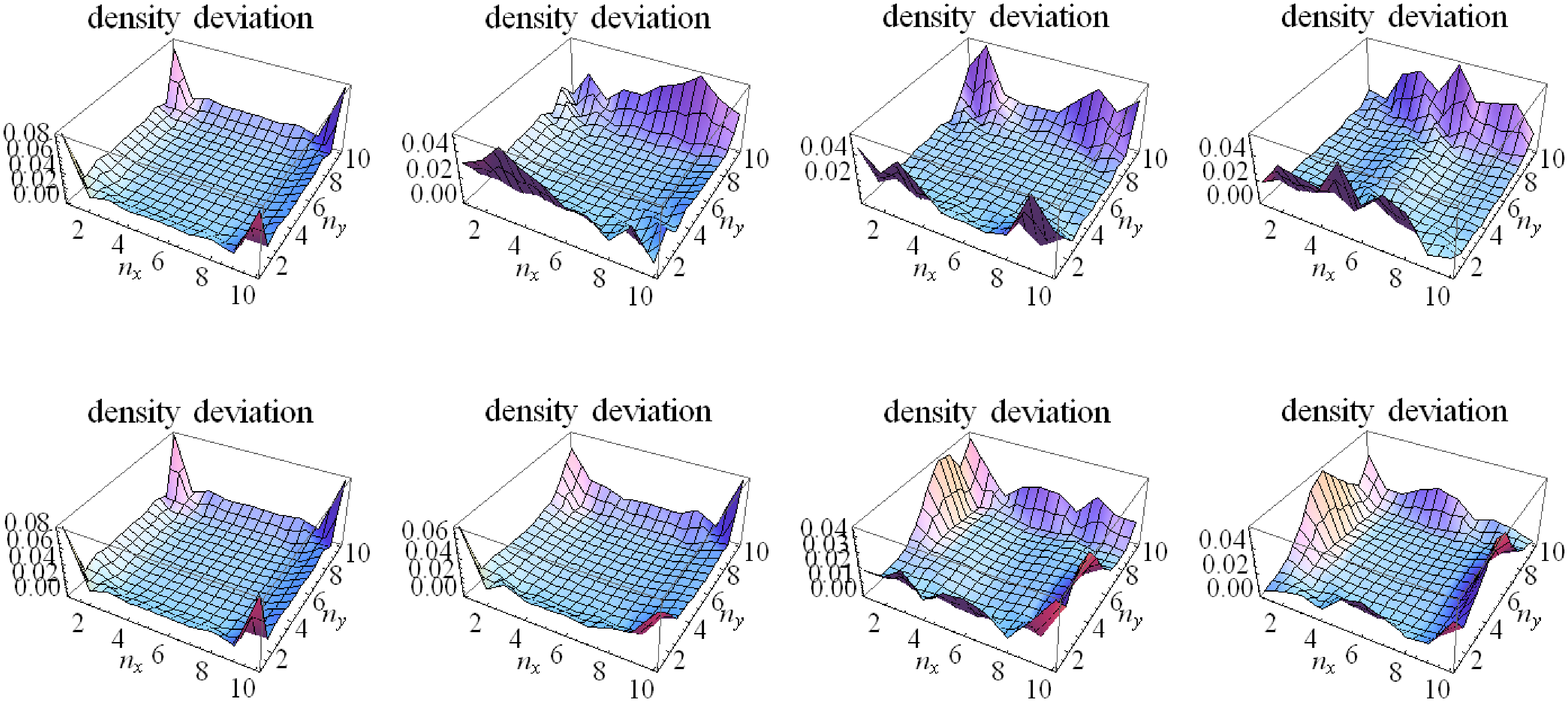}
\caption{Snapshots at $t/t_0=1,8,16,20$ (from left to right) of the density deviation from the average density of the 2D two-band CI model. The top and bottom rows show the cases evolving from OBC to a cylinder and from OBC to a Mobius strip, respectively. We assume that $t_q/t_0=10$, $w=1$, $m=0.5$, and $N=10$. The initial state is the half-filled ground state of the initial Hamiltonian with the edge states occupied.}
\label{CI-r}
\end{figure*}

\subsection{Dynamics of 2D Chern insulator}
\subsubsection{From cylinder or Mobius strip to OBC}
To contrast the features shown in the dynamics of the TQI, we also study time evolution of the 2D two-band Chern insulator (CI) model described by Eq.~(\ref{eq-CI}). In the top row of Figure \ref{CI}, we show the density deviation from the average density of the CI evolving from a cylinder to OBC at time $t/t_0=1,8,16,20$. Here we assume $w=1$, $m=0.5$, and $t_q=10$. Similar to the TQI case, the system size is $N=10$ to allow periodic or twisted boundary condition and ensure the edge modes decays to negligible values away from the open boundary in the static case with OBC. The initial state is the half-filled ground state with the edge states occupied.

For the case from a cylinder to OBC shown in the top row of Figure \ref{CI}, positive density deviation can be observed at the two edges at $n_y=1$ and $n_y=N$ before $t_q$ is reached. This is because the cylinder has open boundary condition along the $y$-axis, so there are two chiral edge modes localized at the two ends of the cylinder. One can see there is density accumulation at the four corners in the final configuration with OBC. However, the density deviation at the corners of the 2D two-band CI model is always positive. Therefore, no quadrupole moment emerges in the CI model, in contrast to the TQI model supporting a quadrupole moment. After the configuration transforms completely to OBC, the two edge states also occupies the sites around $n_x=1$ and $n_x=10$ due to the open boundary along the $x$ direction. Moreover, there are density ripples along both edges in the $x$- and $y$- directions.

The bottom row of Figure \ref{CI} shows the density deviation of the CI model evolving from a Mobius strip to OBC. We choose the same parameters and time slices as those shown in the top row. As shown in Ref.~\cite{Mobius11}, when the 2D two-band CI model is placed on a Mobius strip, the edge states do not cancel completely along a line crossing the bulk. Similar bulk-crossing behavior is observable in our initial condition as there are two positive density peaks located at $n_x=0$ and $n_x=N$, instead of a pair of edge modes along the $x$-direction in the cylindrical configuration. Despite the prominent difference in the initial conditions, the evolution of the density deviation of the case from a Mobius strip to OBC is qualitatively similar to the case from a cylinder to OBC, as the long-time profiles of both cases show four peaks at the corners with density ripples oscillating along both edges.

Similar to the dynamics of the TQI model, we also found memory effects of the ramping rate in the occupation of the topological edge states in the 2D CI model. There are two in-gap chiral edge states and their behavior is similar. Thus, we pick one of them and  shows the time dependence of the occupation of the selected edge state with different ramping time $t_q$ in Figure~\ref{fig:CI_edge}. In absence of dissipation, the occupation of the edge state exhibits steady-state behavior after the transformation of the real-space configuration is completed, despite the whole system does not exhibit steady-state behavior due to the density ripples at the edges. By comparing the steady-state values of the edge-state occupation, one can see that the steady-state value increases with $t_q$ in the case from a cylinder to OBC, as shown in Fig.~\ref{fig:CI_edge}(a). In contrast, the dependence of the steady-state value on $t_q$ is non-monotonic in the case from a Mobius strip to OBC. For instance, the steady-state value of $t_q/t_0=5$ lies above that of $t_q/t_0=10$, as shown in Fig.~\ref{fig:CI_edge}(b), but they are both below the steady-state value of $t_q/t_0=15$.

Importantly, both types of transformations shown in Fig.~\ref{fig:CI_edge} demonstrate that the steady-state occupation explicitly depends on the ramping rate. This is because the edge states in the final OBC configuration are localized states at the boundary, so their population depends on how many particles overlap with the localized states during the evolution, but this in turn depends on how fast the energy levels are changed by the real-space transformation. Since the localized states are the eignestates of the final Hamiltonian but not the intermediate Hamiltonians during the transformation, their population changes with time. In the 2D CI shown in Fig.~\ref{fig:CI_edge}, the population can even exhibit a dip during the real-space transformation. One possible explanation of the dips in Fig.~\ref{fig:CI_edge} is that there are zero-energy eigenstates of the CI as the boundary links change, as shown in Fig.~\ref{fig:TQI_CI_band} in the Appendix. Those zero-energy states do not accumulate dynamic phase and compete with the final localized states to retain excitations. There is no zero-energy state in the TQI, so there is no such competition. Nevertheless, we focus on the steady-state behavior after the transformation is complete.

\subsubsection{From OBC to cylinder or Mobius strip}
Next, we consider the transformation of the real-space configuration from OBC to a cylinder or Mobius strip. Explicitly, $w_{\vn}(t)=w t/t_q$ for $0\le t\le t_q$ and $w_{\vn}=w$ for $t>t_q$ at the $x$-direction boundary. Here we assume $w=1$, $m=0.5$, and $t_q=10$. The system size is $N=10$ and the initial state is the half-filled ground state with the edge state occupied. The top row of Figure~\ref{CI-r} shows the density deviation from the average density of the 2D two-band CI model at time $t/t_0=1,8,16,20$ as the system transforms from OBC to a cylinder with $t_q/t_0=10$. The initial state has four density peaks at the corners with decaying density along the four edges due to the two pairs of chiral edge states residing at the open boundaries along the $x$- and $y$- directions. As the configuration transforms to a cylinder with periodic boundary condition along the $x$-direction, there are density ripples oscillating along the $x$-direction at the two edges of the cylinder. The oscillations do not subside because there is no dissipation in the system.

The bottom row of Figure~\ref{CI-r} shows the density deviation from the average density of the 2D two-band CI model from OBC to a Mobius strip with the same parameters. In this case, there are density ripples both along the edge of the Mobius strip (the $x$-direction in Fig.~\ref{CI-r}) and along a line perpendicular to the edge (the $y$-direction), with the oscillations along the $y$-direction having a larger amplitude. The contrast between the two cases of the CI model is thus similar to the corresponding cases of the TQI model, but the distinction of the directions of the density ripples is more prominent in the TQI case.

We emphasize that the possibility of using the directions of the density ripples to differentiate the underlying configurations is not available in 1D systems because the density ripples spread out to the whole systems, as shown in Ref.~\cite{He2016a}. Thus, more interesting physics can be explored in higher-dimensional topological systems with real-space topological configurations as illustrated here.

\section{Experimental implication}\label{sec:exp}
The original proposal of the TQI~\cite{Bernevig} envisioned its realization in quantum simulators such as cold-atoms in optical potentials. Given the broad choices of quantum systems which may realize various TIs and HOTIs~\cite{Mittal19,Matlack18,Garcia18,Huo19,LinSonic19,Imhof18}, the dynamics analyzed in this work may be studied in systems allowing the formation of a cylinder or Mobius strip. For example, topological systems in a Mobius configuration have been demonstrated by using electronic circuits to simulate quantum particles in lattice potentials~\cite{MobiusSim15}. One may study the dynamics of topological systems by introducing time-dependent parameters in those simulators. Since the phenomena presented here are observable within a few tens of $t_0$, they are feasible even for simulators which may not have a long hold time.

The study here did not include dissipation from system-environment coupling and we assume the particle number is conserved. It is possible to model the time evolution using the quantum master equation~\cite{Breuer_book,Weiss_book} which includes the coupling to the environment or particle exchange. There have been proposals on topological properties of open quantum systems described by the quantum master equation~\cite{Diehl2011b,Bardyn12,Asorey19}, but a systematic study combining both topology of band structure and topology in real space awaits future investigations.

\section{Conclusion}\label{sec:conclusion}
We have shown the transformations of real-space topology can lead to interesting boundary-induced dynamics of topological quantum systems by using the 2D TQI and CI models as examples. From a cylinder or Mobius strip to OBC, the occupation of the quadrupole corner state of the TQI and the edge state of the CI reaches steady states after the transformation. The steady-state values depend on the ramping rate of the transformation and exhibit quantum memory effect. In addition, the physical quadropole moment of the TQI exhibits similar quantum memory effect. On the other hand, the systems develop density ripples instead of steady states as they transform from OBC to a cylinder or Mobius strip. While the density ripples are confined in the edges of a cylinder, they traverse the bulk in the case of a Mobius strip and reflect the real-space topology. Generalizations of the boundary-induced dynamics to higher-dimensional systems or interacting systems may reveal rich physics of topological systems out of equilibrium.

\begin{acknowledgements}
We thank Prof. Hao Guo for stimulating discussions.	Y. H. is supported by NSFC under Grant No. 11874272.
\end{acknowledgements}

\appendix
\section{Details of time evolution and initial condition}

\subsection{Details of time evolution of 2D TQI}\label{sec:Evo_TQI}
The time evolution equation of the fermion operator is
\be
&&\frac{d \psi_{\vn}}{d t}=-i\Big[\Big(m(\Gamma_2+\Gamma_4)+\delta\Gamma_0\Big)\psi_{\vn}\nonumber\\
&&+\frac{w_{\vn}}2(\Gamma_4-i\Gamma_3)\psi_{\vn+\hx}+\frac{w_{\vn-\hx}}2(\Gamma_4+i\Gamma_3)\psi_{\vn-\hx}\nonumber\\
&&\quad+\frac{w}2(\Gamma_2-i\Gamma_1)\psi_{\vn+\hy}+\frac{w}2(\Gamma_2+i\Gamma_1)\psi_{\vn-\hy}\Big].
\ee
In components, the above equation can be written as
\be
&&i\frac{dc_{\vn,1}}{dt}=\Big[w_{\vn-\hx} c_{\vn-\hx,3}-w c_{\vn+\hy,4}
+m(c_{\vn,3}-c_{\vn,4})+\delta c_{\vn,1}\Big],\nonumber\\
&&i\frac{dc_{\vn,2}}{dt}=\Big[w c_{\vn-\hy,3}+w_{\vn} c_{\vn+\hx,4}
+m(c_{\vn,3}+c_{\vn,4})+\delta c_{\vn,2}\Big],\nonumber\\
&&i\frac{dc_{\vn,3}}{dt}=\Big[w_{\vn} c_{\vn+\hx,1}+w c_{\vn+\hy,2}
+m(c_{\vn,1}+c_{\vn,2})-\delta c_{\vn,3}\Big],\nonumber\\
&&i\frac{dc_{\vn,4}}{dt}=\Big[w_{\vn-\hx} c_{\vn-\hx,2}-w c_{\vn-\hy,1}
-m(c_{\vn,1}-c_{\vn,2})-\delta c_{\vn,4}\Big].\nonumber
\ee

The time-evolution of the correlation function~\eqref{eq:Gnm} is given by
\begin{widetext}
	\be
	\frac{d}{dt}G(\vn,\vm)
	&=&-i\Big[\frac{w_{\vn}}2(\Gamma_4-i\Gamma_3)G(\vn+\hx,\vm)
	+\frac{w_{\vn-\hx}}2(\Gamma_4+i\Gamma_3)G(\vn-\hx,\vm)
	+\frac{w}2(\Gamma_2-i\Gamma_1)G(\vn+\hy,\vm) \nonumber\\
	&&+\frac{w}2(\Gamma_2+i\Gamma_1)G(\vn-\hy,\vm)
	+\Big(m(\Gamma_2+\Gamma_4)+\delta\Gamma_0\Big)G(\vn,\vm)\Big]
	+i\Big[\frac{w_{\vm}}2G(\vn,\vm+\hx)(\Gamma_4+i\Gamma_3) \nonumber\\
	&&+\frac{w_{\vm-\hx}}2G(\vn,\vm-\hx)(\Gamma_4-i\Gamma_3)
	+\frac{w}2G(\vn,\vm+\hy)(\Gamma_2+i\Gamma_1)+\frac{w}2G(\vn,\vm-\hy)(\Gamma_2-i\Gamma_1) \nonumber\\
	&&+G(\vn,\vm)\Big(m(\Gamma_2+\Gamma_4)+\delta\Gamma_0\Big)\Big].
	\ee
	In the expression, matrix multiplications are implicitly assumed.
	In components, we have
	\be
	&&\frac{d}{dt}G_{11}(\vn,\vm)=
	-i\Big[w_{\vn-\hx}G_{31}(\vn-\hx,\vm)-w G_{41}(\vn+\hy,\vm)
	+m G_{31}(\vn,\vm)-m G_{41}(\vn,\vm)+\delta G_{11}(\vn,\vm)\Big]\nonumber\\
	&&+i\Big[w_{\vm-\hx}G_{13}(\vn,\vm-\hx)-w G_{14}(\vn,\vm+\hy)
	+m G_{13}(\vn,\vm)-m G_{14}(\vn,\vm)+\delta G_{11}(\vn,\vm)\Big],\nonumber\\
	&&\frac{d}{dt}G_{21}(\vn,\vm)=
	-i\Big[w G_{31}(\vn-\hy,\vm)+w_{\vn} G_{41}(\vn+\hx,\vm)
	+m G_{31}(\vn,\vm)+m G_{41}(\vn,\vm)+\delta G_{21}(\vn,\vm)\Big]\nonumber\\
	&&+i\Big[w_{\vm-\hx}G_{23}(\vn,\vm-\hx)-w G_{24}(\vn,\vm+\hy)
	+m G_{23}(\vn,\vm)-m G_{24}(\vn,\vm)+\delta G_{21}(\vn,\vm)\Big],\nonumber\\
	&&\frac{d}{dt}G_{31}(\vn,\vm)=
	-i\Big[w_{\vn}G_{11}(\vn+\hx,\vm)+w G_{21}(\vn+\hy,\vm)
	+m G_{11}(\vn,\vm)+m G_{21}(\vn,\vm)-\delta G_{31}(\vn,\vm)\Big]\nonumber\\
	&&+i\Big[w_{\vm-\hx}G_{33}(\vn,\vm-\hx)-w G_{34}(\vn,\vm+\hy)
	+m G_{33}(\vn,\vm)-m G_{34}(\vn,\vm)+\delta G_{31}(\vn,\vm)\Big],\nonumber\\
	&&\frac{d}{dt}G_{41}(\vn,\vm)=
	-i\Big[w G_{11}(\vn-\hy,\vm)+w_{\vn-\hx}G_{21}(\vn-\hx,\vm)
	-m G_{11}(\vn,\vm)+m G_{21}(\vn,\vm)-\delta G_{41}(\vn,\vm)\Big]\nonumber\\
	&&+i\Big[w_{\vm-\hx}G_{43}(\vn,\vm-\hx)-w G_{44}(\vn,\vm+\hy)
	+m G_{43}(\vn,\vm)-m G_{44}(\vn,\vm)+\delta G_{41}(\vn,\vm)\Big].\nonumber
	\ee
\end{widetext}
There are another 12 equations for $G_{ab}$ with $a=1,2,3,4$ and $b=2,3,4$.

\subsection{Initial condition of 2D TQI}\label{sec:Ini_TQI}
\subsubsection{Cylinder and OBC configurations}
For the Hamiltonian~(\ref{Hxy}) with open boundary condition along the $y$ axis and periodic boundary condition along the $x$ axis, we use the initial condition of the half-filled ground state of fermions. In the first quantized form, the Hamiltonian can be written as
\be
&&H_1=[m(\Gamma_2+\Gamma_4)+\delta\Gamma_0]\otimes I_0\otimes I_0\nonumber\\
&&+\frac12(\Gamma_4-i\Gamma_3)\otimes T_1\otimes I_0
+\frac12(\Gamma_2-i\Gamma_1)\otimes I_0\otimes T_2\nonumber\\
&&+\frac12(\Gamma_4+i\Gamma_3)\otimes T_1^T\otimes I_0
+\frac12(\Gamma_2+i\Gamma_1)\otimes I_0\otimes T_2^T.\nonumber\\
\label{H1}
\ee
Here we introduce $N$ by $N$ matrices $I_0=\delta_{ij}$, $T_1=\delta_{i,i+1}+\delta_{N,1}$ and $T_2=\delta_{i,i+1}$. $T_1^T$ and $T_2^T$ are the transpose of $T_1$ and $T_2$. Then, we numerically diagonalize $H_1$ to find the $m$-th eigenvector $\phi_{im}$ corresponding to the eigenvalue $E_m$. Here $i=1,\cdots,4N^2$ labels the components. The labels of fermion operators in Eq.~(\ref{H1}) and Eq.~(\ref{Hxy}) are related by $c_{N^2(a-1)+N(n_x-1)+n_y}=c_{a,n_x,n_y}$. Here $a=1,2,3,4$ is the orbital index of fermions. We introduce the quasi-particle operator $\eta_m=\sum_i\phi^*_{im}c_i$, where $*$ denotes complex conjugate. The Hamiltonian becomes
\be
H_1=\sum_m E_m\eta^{\dagger}_m\eta_m.
\ee
The initial state is $\ket{S_0}=\prod_{E_m \le 0}\eta^{\dagger}_m\ket{0}$. The initial correlation functions can be obtained from
\be
\ep{c_i\dc_j}=\sum_{E_m>0}\phi_{im}\phi^*_{jm}.
\ee

For the case with OBC, the Hamiltonian is given by Eq.~(\ref{H1}) except $T_1=\delta_{i,i+1}$. The replacement reflects the corresponding boundary condition. The initial condition is then obtained by the same procedure as those for the cylindrical case.

\subsubsection{Mobius strip configuration}
With twisted boundary condition along the $x$ axis and OBC along the $y$ axis, the fermion operator satisfies $\psi_{N,n_y}=\psi_{1,N-n_y+1}$ and the system resembles a Mobius strip. In the first quantized form, the Hamiltonian can be written as
\be
&&H_2=[m(\Gamma_2+\Gamma_4)+\delta\Gamma_0]\otimes I_0\otimes I_0\nonumber\\
&&+\frac12(\Gamma_4-i\Gamma_3)\otimes T_2\otimes I_0
+\frac12(\Gamma_2-i\Gamma_1)\otimes I_0\otimes T_2\nonumber\\
&&+\frac12(\Gamma_4+i\Gamma_3)\otimes T_1^T\otimes I_0
+\frac12(\Gamma_2+i\Gamma_1)\otimes I_0\otimes T_2^T\nonumber\\
&&+\frac12(\Gamma_4-i\Gamma_3)\otimes T_3\otimes T_4
+\frac12(\Gamma_4+i\Gamma_3)\otimes T_3^T\otimes T_4^T.\nonumber\\
\label{H2}
\ee
Here $T_3=\delta_{N,1}$ and $T_4=\delta_{i,N-i+1}$. We numerically diagonalize $H_2$ to find the $m$-th eigenvector $\phi_{im}$ corresponding to the eigenvalue $E_m$. The quasi-particle operator is defined as $\eta_m=\sum_i\phi^*_{im}c_i$. The initial state is then given by $\ket{S_0}=\prod_{E_m \le 0}\eta^{\dagger}_m\ket{0}$, and the initial correlation function can be obtained from
\be
\ep{c_i\dc_j}=\sum_{E_m>0}\phi_{im}\phi^*_{jm}.
\ee

\subsection{Details of time evolution and initial condition of 2D CI}\label{sec:Evo_CI}
The time-evolution equation for $\psi_{\vn}$ is given by
\be
&&\frac{d\psi_{\vn}}{dt}=-i\sum_{\vn}\Big[m\sigma_3\psi_{\vn}\nonumber\\
&&+\frac{w_n}{2}(\sigma_3-i\sigma_1)\psi_{\vn+\hx}
+\frac{w}{2}(\sigma_3-i\sigma_2)\psi_{\vn+\hy}\nonumber\\
&&+\frac{w_{\vn-\hx}}{2}(\sigma_3+i\sigma_1)\psi_{\vn-\hx}
+\frac{w}{2}(\sigma_3+i\sigma_2)\psi_{\vn-\hy}
\Big].
\ee
The time-evolution equation of correlation function is
\be
&&i\frac{d}{dt}G(\vn,\vm)=\Big[m\sigma_3 G(\vn,\vm)-G(\vn,\vm)m\sigma_3\Big]\nonumber\\
&&+\Big[\frac{w_{\vn}}2\sigma_{31}G(\vn+\hx,\vm)
+\frac{w_{\vn-\hx}}2\sigma_{31}^{\dagger}G(\vn-\hx,\vm)\nonumber\\
&&+\frac{w}2\sigma_{32}G(\vn+\hy,\vm)+\frac{w}2\sigma_{31}^{\dagger}G(\vn-\hy,\vm)\Big]\nonumber\\
&&-\Big[\frac{w_{\vm}}2G(\vn,\vm+\hx)\sigma_{31}^{\dagger}
+\frac{w_{\vm-\hx}}2G(\vn,\vm-\hx)\sigma_{31}\nonumber\\
&&+\frac{w}2G(\vn,\vm+\hy)\sigma_{32}^{\dagger}+\frac{w}2G(\vn,\vm-\hy)\sigma_{32}\Big].
\ee
Here we have defined $\sigma_{31}=\sigma_3-i\sigma_1$ and $\sigma_{32}=\sigma_3-i\sigma_2$.

The initial state can be obtained by the following Hamiltonian in the first quantized form:
\be
&&H_{CI1}=m\sigma_3\otimes I_0\otimes I_0\nonumber\\
&&+\frac12(\sigma_3-i\sigma_1)\otimes T_1\otimes I_0
+\frac12(\sigma_3-i\sigma_2)\otimes I_0\otimes T_2\nonumber\\
&&+\frac12(\sigma_3+i\sigma_1)\otimes T_1^T\otimes I_0
+\frac12(\sigma_3+i\sigma_2)\otimes I_0\otimes T_2^T.\nonumber\\
\ee
Here we introduce $N$ by $N$ matrices $I_0=\delta_{ij}$, $T_1=\delta_{i,i+1}+\delta_{N,1}$, and $T_2=\delta_{i,i+1}$. $T_1^T$ and $T_2^T$ are the transpose of $T_1$ and $T_2$. By imposing boundary condition corresponding to the configurations (1), (2), or (3), we numerically diagonalize $H_{CI1}$ to find the $m$-th eigenvector $\phi^{CI}_{im}$ corresponding to the eigenvalue $E^{CI}_m$. The quasi-particle operator is given by $\eta^{CI}_m=\sum_i(\phi^{CI}_{im})^*c_i$, and the initial state is $\ket{S_0}=\prod_{E^{CI}_m \le 0}(\eta^{CI}_m)^{\dagger}\ket{0}$. The initial correlation function can be obtained from
\be
\ep{c_i\dc_j}=\sum_{E^{CI}_m>0}\phi^{CI}_{im}(\phi^{CI}_{jm})^*.
\ee

\begin{figure}
	\centerline{\includegraphics[width=\columnwidth]{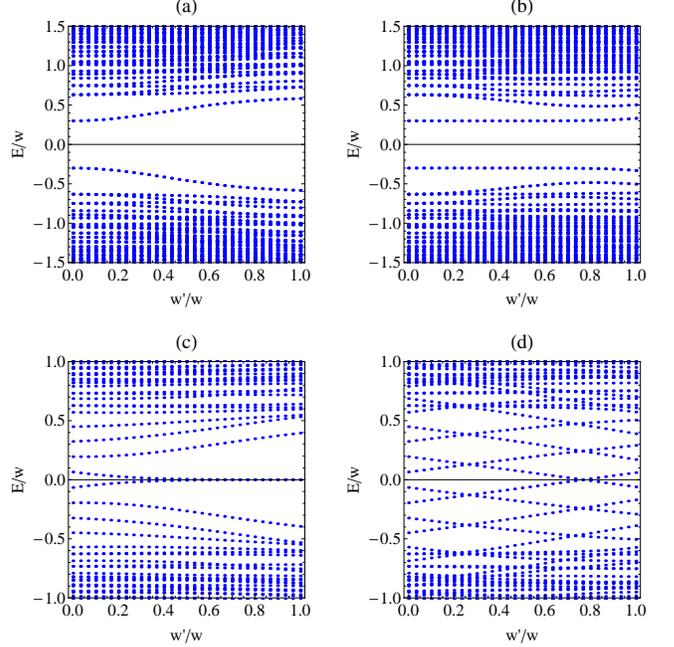}}
	\caption{Eigenenergy spectra of the 2D TQI [(a) and (b)] and CI [(c) and (d)] as a function of $w'/w$ on a $10\times10$ lattice. Here $w'$ is the hopping coefficient at the boundary in the $x$-direction. The systems are open in the $y$-direction. When $w'/w=1$, the systems have a cylinder configuration for (a) and (c) and a Mobius-strip configuration for (b) and (d). When $w'/w=0$, the systems have open boundary condition in both directions. Here $w=1$ and $m=0.5$ for all panels and $\delta=0.2$ for (a) and (b).}
	\label{fig:TQI_CI_band}
\end{figure}

\subsection{Eigenenergy spectra of TQI and CI}
The eigenenergy spectra of the 2D TQI and CI on a $10\times 10$ square lattice with open boundary condition in the $y$-direction and a varying hopping coefficient $w'$ along a line in the $x$-direction are shown in Fig.~\ref{fig:TQI_CI_band}. We considered both periodic and twisted structures along the $x$-direction, so the systems become a cylinder (or a Mobius strip) when $w'/w=1$ in panels (a) and (c) (or (b) and (d)) of Fig.~\ref{fig:TQI_CI_band}. However, $w'/w=0$ leads to the same OBC configuration for all the cases. In the thermodynamic limit, the spectra will become the band structures of the corresponding systems.

The localized edge states are the in-gap states near $E=0$. Due to the open-boundary condition along the $y$-direction in the 2D cases, the edge states are always present in the eigenenergy spectra, but their energies depend on $w'/w$. This feature is different from the 1D cases shown in Ref.~\cite{He2016a}, where the edge states only become the genuine eigenstates in the fully open-boundary case. We remark that Fig.~\ref{fig:TQI_CI_band} only depicts the eigenenergies of the Hamiltonian with the corresponding $w'/w$. The dynamics induced by a change of the boundary condition generates particle excitations jumping among the eigenstates, leading to the results shown in the main text.

%

\end{document}